\documentstyle[11pt]{article}
\newcommand{\blankline}{\vskip .3cm}
\newcommand{\f}{\begin{equation}}
\newcommand{\ff}{\end{equation}}
\begin{document}
\vfill
\centerline{\Large \bf Outline}
\vskip .1cm
\centerline{\Large \bf of a Generally Covariant Quantum Field
Theory}
\vskip .1cm
\centerline{\Large \bf and a Quantum Theory of Gravity}
\rm
\vskip.7cm
\centerline{\bf Carlo Rovelli}
\vskip .3cm
\centerline{\it Department of Physics, University of
        Pittsburgh, Pa 15260, USA}
\centerline{\it E-mail: rovelli@vms.cis.pitt.edu}
\vskip .3cm\vskip .3cm
\vfill
\centerline{\today}
\vfill
\blankline
\centerline{Abstract}
\noindent
\blankline
We study a tentative generally covariant quantum field theory,
denoted the T-Theory, as a tool to investigate the consistency
of quantum general relativity.   The theory describes the gravitational
field  and a minimally coupled scalar field; it is based on the loop
representation, and on a certain number of quantization choices.
Four-dimensional diffeomorphism-invariant quantum transition
probabilities can be computed from the theory.  We present the
explicit calculation of the  transition  probability between two
volume eigenstates as an example.  We  discuss the  choices on
which the T-theory relies, and the possibilities of  modifying them.
 \vfill
\thispagestyle{empty}
\eject
\section{Introduction}

In this work we construct a generally covariant quantum field
theory, based
on the classical theory of the gravitational field interacting with a
minimally
coupled scalar field. We illustrate how quantum transition
amplitudes can be
computed from this theory without breaking four-dimensional
diffeomorphism invariance, and we compute some of these
transition
amplitudes explicitly.   Our first motivation is to explore
diffeomorphism invariant
quantum field theories with infinite degrees of freedom, in view
of the quantum gravity puzzle \cite{cris}.
Diffeomorphism invariance has far reaching consequences:
since no
quantity local in the space-time coordinates can be diffeomorphism
invariant,
a genuinely general covariant quantum field theory cannot be
constructed in
the conventional framework of local quantum field theory, as it is
synthesized for instance in \cite{haag}.  The physical meaning of
diffeomorphism invariance consists in the fact that locality is only
defined with respect to dynamical objects of the theory itself -- we
think that this idea captures General Relativity's main discovery
about Nature.  The theory constructed in this paper is an attempt to
incorporate this idea within a non-trivial quantum field theory.

The second motivation of this work is to carry on the development of
the loop representation approach to quantum general relativity
\cite{carlolee}. This research program has advanced in two main
directions during the last few years.  On the one hand, the credibility
of the results obtained has been strengthened by the development of
a
mathematical-physics approach that has put these results on a firm
ground \cite{abhayetal}.   On the other hand, a number of novel
results \cite{weaves,fermions,gravitons,area,volume,hamiltonian}
and calculation tools \cite{spinnet} have moved the theory forward,
to the point where one can begin to perform physical calculations
and to address the issues of the theory's consistency and physical
implications.  In its present form, however, the loop representation
of quantum gravity is not a complete theory
 \cite{report,bernd,ab};
essential missing elements are the complete determination of the
scalar product, the choice of the ordering of certain key operators,
and a general recipe for defining and computing diffeomorphism
invariant physical expectation values.   Various ideas and various
proposals for solving each of these problems have been put forward,
but it is not clear whether there exists
a combination of those proposals that
yields a consistent quantum theory with the correct classical
limit.  The discussion on problems such as ordering, definition of
diffeomorphism invariant quantum observables, or choice of scalar
product, has traditionally been quite academic and ideas were tested
only within oversimplified finite dimensional models.   The recent
developments of the loop representation that we mentioned allow
ideas to be tested in the realistic case.  In this paper, we complete
the definition of the loop representation theory by choosing a
number of assumptions, listed below.  Our intent is explorative
\cite{einstein} and we do not take these assumptions for granted in
any sense: the theory defined by these assumptions is a ``tentative''
theory, likely to turn out either inconsistent or physically
incorrect;  accordingly, we denote it as the
Tentative-theory, or T-theory.   We present this theory as a possible
candidate for a non trivial generally covariant quantum field theory
and in order to begin exploring the possible ways of completing the
loop representation.

The main ingredients of our construction are the following.  We
begin with the classical theory formed by General Relativity with
cosmological constant term and a minimally coupled scalar
matter field $\phi(x)$.  Following \cite{leebrill} and \cite{karel},
we study this
theory in the gauge $\phi(\vec x,t)\propto t$. In this gauge
the local degree of freedom of the scalar field disappears and
reappears as a new ``longitudinal'' degree of freedom of the
gravitational field, a phenomenon analogous to the disappearance of
the Higgs field that gives mass to gauge bosons.  In the gauge fixed
form, the theory has a genuine Hamiltonian (instead of a
hamiltonian constraint as the non-gauge-fixed theory), which
generates the evolution of the gravitational
field as a function of the value of the spatially constant scalar field.
In other words, we regard the scalar field as the independent
variable for the temporal localization of events.   We quantize this
theory using the loop representation \cite{gambini} and we make use
of the result of reference \cite{hamiltonian}, where the
loop-representation hamiltonian operator was
constructed and shown to be finite.  We use  the recently discovered
spin network basis \cite{spinnet}. This basis, which was suggested
by  T. Thiemann \cite{thomas} and independently discovered by J.
Baez \cite{baezsn} and T. Foxon \cite{foxon}
(and perhaps others), is the eigenbasis of the
volume operator
\cite{volume}.  We assume that this basis is orthonormal.    This
assumption fixes the scalar product.  The computation of the action
of the Hamiltonian on the spin network basis states has been
recently completed \cite{roumen}.
The expression for the Hamiltonian contains the square root of a
certain
operator.  The key technical step we take here is the
explicit computation of this square root.  This yields an explicit
expression for the action of the Hamiltonian in the spin network
basis.  Then we can use perturbation theory, as first
suggested by L. Smolin \cite{ls}, to compute first order
transition amplitudes between volume eigenstates.   In particular, in
this paper we shall compute probabilities, according to the
T-theory, that if the system is in a
volume eigenstate $|s_i>$ when the scalar field has value $t$, it be
later
found in a volume eigenstate $|s_f>$ when the scalar field has
value
$t+\Delta t$.

Let us list here the assumptions that define the T-theory. We
will discuss them, as well as the possibility of modifying them, in
the
conclusion.
\begin{itemize}
\item Choice of the scalar product: The Hilbert space structure is
defined by
the orthonormality of the spin network basis.
\item  Evolution:  It make sense to deal with the temporal part of
 diffeomorphism invariance by considering quantum evolution
with
respect to one of the dynamical variables --the scalar field. The
scalar field
can loosely be interpreted as a clock field.  We refer to the abundant
literature on the so called ``Issue of Time'' in quantum gravity for
the numerous alternative positions on this matter \cite{christime}.
Here, we follow essentially \cite{carlotime}.
\item An essential aspect of the theory below is the presence of
the cosmological constant.  Physically, it is needed to
balance the energy density of the clock field, so that the clock can
run fast without crumpling the universe.
\item Locality: No notion of locality is given in the theory in general.
Quantum states are represented by abstract combinatorial and
topological
relations and have no space-time localization whatsoever. Spatial
localization
emerges only approximately and is defined with respect to the
quantum
state itself \cite{weaves}.
\item The ordering of the Hamiltonian is the one chosen in the
references \cite{hamiltonian,roumen}.
\end{itemize}
The main message of the present paper is that the loop
representation of General Relativity,
supplemented by these assumptions yields a
quantum theory of gravity, the T-theory, in which physical
transitions amplitudes can be computed.  Thus, diffeomorphism
invariant quantum field theoretical calculations can be performed.

This paper is organized as follows.  Section 2 presents the classical
theory.  Section 3 describes the quantum theory and the Hamiltonian.
In section 4 the square root is computed. In section 5 perturbation
theory is introduced, and a transition amplitude is explicitly
computed. In section 6 we discuss the assumptions on which the
theory is based, present our conclusions and indicate
future directions of research.

\section{Classical theory}

We consider General Relativity, with a cosmological constant term,
and a minimally coupled massless scalar field $\phi(x)$. The action
is
\f
	S[g_{\mu\nu},\phi] = - \int d^4x \sqrt{g}
	\left\{
		{R\over 16\pi G}
		- {\lambda\over 16\pi G}
		+ {1\over 2}\,
		g^{\mu\nu}\partial_\mu\phi \partial_\nu\phi
	\right\}.
\label{action}
\ff
We follow \cite{weinberg} for notation and conventions.  We put the
velocity of light equal to one, but we indicate the Planck constant
$\hbar$ and the Newton constant $G$ explicitly.  We assume that
the constant $\lambda$ is positive.  With the signs in (\ref{action}),
this choice corresponds to a negative cosmological energy density.
The equations of motion for the gravitational field are
\f
	R_{\mu\nu}-{1\over 2}Rg_{\mu\nu}
	+{1\over 2} \lambda g_{\mu\nu} =
	- 8\pi G
\left(\partial_\mu\phi \partial_\nu\phi
	- {1\over 2}g_{\mu\nu} \partial^\rho\phi \partial_\rho\phi
\right).
	\label{eqmot}
\ff
We partially fix the gauge invariance of the theory as follows. We
choose the time coordinate $t$ according to
\f
	\phi(\vec x, t) = \mu t.
	\label{gauge}
\ff
The constant $\mu$ has been inserted in order to adjust dimensions.
Later we will choose a particularly convenient value for $\mu$.
The scalar field has dimensions of $\sqrt{M/L}$ (M is mass and L is
length, we have $c=1$). Thus $\mu$ has dimensions $\sqrt{M/L^3}$.
In fixing the gauge (\ref{gauge}), we  restrict the domain of
application of the theory, because in a general solution of
(\ref{eqmot}) the regions of constant $\phi$ may not be
space-like hypersurfaces, as the $t=constant$ surfaces have to be.
Thus, we must restrict the theory to those solutions and those
space-time
regions in which the surfaces $\phi(\vec x, t) = constant$
define a space like foliation of space-time.  Let us denote those
regions as the ``clock regime''.   The classical theory is fully
consistent in the gauge (\ref{gauge}), provided we restrict to this
regime.   The physical interpretation of this gauge is simple: because
of general covariance, General Relativity does not describe evolution
in an external absolute time, but rather it describes the relative
evolution of gravitational and matter variables  with respect to each
other.  We thus pick the scalar field $\phi$, rather arbitrarily, as the
independent physical variable with respect to which we consider the
evolution of the other physical variables.   In this very weak sense
we may view the scalar field as a phenomenological description of a
clock.  Later, we will see that in appropriate circumstances the
value of $\phi$ coincides with proper time. As any physical clock in
General
Relativity, the clock $\phi$
ceases to behave as a good clock in certain physical regions.  Since
evolution can bring the gravitational system into these regions, the
evolution equations in the gauge (\ref{gauge}) may blow up.   Such a
blow up signals the exit from the domain of validity of the gauge
fixed formalism:  from the physical point of view it signals the fact
that the object we are taking as clock has ceased to behave as a
good clock.  In particular, it could stay still, or run backwards.  The
first of the assumptions on which the quantum theory we are going
to build relies is that this procedure, which is fully viable in the
classical theory, is also viable (within the approximation we are
going to take) in the quantum theory.

Gauges of the form (\ref{gauge}) are well known in General
Relativity, and commonly considered in discussing the intrinsic
observable evolution; indeed, the idea of using dynamical matter to
define preferred coordinates goes back to Einstein's papers. See for
instance \cite{tradition}.  In particular the use of a scalar field was
considered in \cite{karel}, where advantages and difficulties of this
choice are discussed.  Our treatment is almost identical to
the one of Smolin
in \cite{leebrill}, where the gauge $\vec\nabla\phi=0$ was
considered. For clarity, we discuss here the relation between the
two gauges.  The gauge choice considered by Smolin is time
independent, and therefore well defined on the conventional ADM
phase space.  It reduces the four dimensional diffeomorphism group
down to the product of the three dimensional diffeomorphism group
times the reparametrizations of a single variable, and it fixes all
the infinite degrees of freedom of the scalar field variable except
one: its spatially constant, time dependent value $\Phi(t)=\phi(\vec
x, t)$.
Smolin then makes a ``choice of intrinsic time'', by picking $\Phi$ as
the independent variable with respect to which evolution of the rest
of the variables is considered.  The gauge choice (\ref{gauge})
considered here, on the other hand, is stronger than Smolin's gauge;
it reduces the four dimensional diffeomorphism group fully down to
the three dimensional one.  Since there is no residual time
reparametrization invariance in the gauge fixed theory, the theory
has a genuine time evolution, and the (controversial?) problem of
``choosing an intrinsic time'' within the canonical scheme does not
appear.  The results, at the end of the day, are equivalent.

As a first step towards the quantum theory, we derive the
hamiltonian formalism of the theory in the gauge (\ref{gauge}). We
follow the traditional non-covariant derivation, but in the present
context it is
worthwhile to recall the well known fact that the result of
such a
derivation, namely the hamiltonian formalism by itself, is fully
covariant (see
for instance \cite{luca}).
The
complete procedure is to first derive the hamiltonian formalism and
then add the gauge fixing; however, the same result is
obtained in the present case by means of the short-cut of directly
inserting the gauge choice into the Lagrangian.  This procedure has
the additional advantage of saving us from the complications of the
time dependent gauge fixing.  If we do so, we
obtain the lagrangian density
\f
	L = - \sqrt{-g}
	\left\{
		{R-\lambda\over 16\pi G}
		+ {\mu^2\over 2}
		g^{00} \right\}.
\ff
The theory is still invariant under spatial diffeomorphisms, but not
anymore under time diffeomorphisms.  Replacing the 4-dimensional
metric variable with the ADM variables $N=(-g^{00})^{-1/2}, \
N^a=g_{0a},\ q_{ab}=g_{ab}$, where $a,b...=1,2,3$, we obtain (using
the
Gauss-Codazzi equations), the ADM form of the lagrangian density
\f
	L = N\sqrt{q}
	\left\{
		{k_{ab} k^{ab}-k^2 \over 16\pi G N^2}
		+ {\lambda-R\over 16\pi G}
		+ {\mu^2\over 2 N^2}
		\right\},
\ff
where
\f
	k_{ab} = {1\over 2} (\dot q_{ab} - D_aN_b - D_bN_a)
\ff
($N^{-1}$ times the extrinsic curvature of the constant-$t$
surfaces), and $R$  --from now on-- is the scalar curvature of
the spatial metric
$q_{ab}$.  Since the equations of motion obtained by varying the
Lapse function $N$ can be solved for $N$ itself, we can insert
the solution back into the Lagrangian, obtaining the
same equations of motion, and there is no constraint
associated with $N$ (See for instance \cite{n}).
The equation that we obtain varying the Lapse is
\f
	N^2 = {k_{ab} k^{ab}-k^2 + \mu^2  8\pi G \over \lambda - R}
\label{N}
\ff
Inserting this value for $N$ back into the lagrangian density, we
obtain
\f
	L = - {\sqrt{q}\over 8\pi G}\sqrt{(\lambda-R)(k_{ab}
	k^{ab}- k^2 + 8\pi G \mu^2)},
\ff
which, as can be verified, yields the correct equations of motion.
The hamiltonian formalism is then easily derived.  The only
hamiltonian constraint is the usual diffeomorphism constraint $C_a
= D_b p^{ab}$, where $p^{ab}$ is the momentum conjugate to the
three metric $q_{ab}$, and a tedious but unproblematic computation
yields the Hamiltonian
\f
	H = \sqrt{2}\mu \int d^3x\ \sqrt{-16\pi G (p_{ab}p^{ab}-p^2)
	+ {\lambda-R\over 16\pi G} q },
	\label{hamiltoniano}
\ff
We emphasize the fact that (\ref{hamiltoniano}) is not a constraint,
but a genuine Hamiltonian.  The correctness of this direct derivation
of the hamiltonian formalism can be verified
by checking the evolution equations that the formalism
determines. These reproduce the Einstein equations in the gauge
(\ref{gauge}).  Thus, the classical hamiltonian theory in the
particular gauge considered is defined by the three metric $q_{ab}$
and its conjugate variable $p^{ab}$, subjected to the conventional
first class ADM diffeomorphism constraint and evolving under the
evolution generated by the Hamiltonian (\ref{hamiltoniano}). This
evolution is to be interpreted as the evolution of the metric variable
as a function of the field variable $\phi$.  The
Hamiltonian (\ref{hamiltoniano}) can be written in term of the
conventional
ADM
hamiltonian constraint (with cosmological constant)
$C^{(\lambda)}_{ADM}$ of the pure gravity theory as
\f
H = \sqrt{2}\mu\int d^3x\ \sqrt{q}\ \sqrt{-C^{(\lambda)}_{ADM}}.
\label{hc}
\ff
Notice that in spite of the fact that only gravitational variables
appear in this gauge, the theory has 3 degrees of freedom per space
point (6 components of $q_{ab}(x)$ minus three first class
constraints $C_a(x)$) which correspond to the 2 degrees of freedom
per point of the gravitational field, plus the degree of freedom of
the scalar field.

One way of understanding the Hamiltonian
(\ref{hc}) is to consider the hamiltonian constraint of the
non-gauge-fixed theory, which is
\f
	q C^{(\lambda)}_{ADM} + {1\over 2} \Pi^2 = 0
\label{pos}
\ff
where $\Pi$ is the momentum conjugate to the scalar field.
In the regions in which the time derivative of the scalar field,
and therefore its conjugate momentum $\Pi$,
are positive definite, this is
equivalent to the constraint
\f
	\Pi + \sqrt{2}\,\sqrt{q}\,\sqrt{-C^{(\lambda)}_{ADM}} = 0.
\label{hhh}
\ff
In particular, we may consider the evolution generated by the
component of the hamiltonian constraint obtained by integrating
(\ref{hhh}) in space
\f
	\mu\int \Pi + \sqrt{2}\mu\int\sqrt{q}\,\sqrt{-
C^{(\lambda)}_{ADM}} = 0.
\ff
The first term evolves only the scalar field, yielding (for suitable
initial data)
$\phi(x,t) = \mu t$, which is our gauge choice; the second term
evolves only the gravitational variables, and is equal to the
Hamiltonian (\ref{hc}) we have derived.   This short re-derivation is
incomplete by itself, but has some virtues. First, it confirms that
every evolution generated by the Hamiltonian (\ref{hc}) is indeed a
solution of Einstein's equation in the coordinates in which $\phi(x,t)
=
\mu t$. Second, it clarifies the origin of the curious $\sqrt{2}$
factor
in (\ref{hc}). Third, and most importantly, it shows that on all
physical solutions the term $-C^{(\lambda)}_{ADM}$ inside the
square root in
the
Hamiltonian is always non-negative.  This is clear from  (\ref{pos}).
Therefore, the exit of the system from the clock regime does not
correspond to the Hamiltonian becoming imaginary, as one might
have imagined, but to the vanishing of the Hamiltonian
density.\footnote{Since the derivative of a square root is its
inverse, the inverse of the square root appears in the denominator of
the Hamilton equations generated by $H$, and thus vanishing of the
square root yields divergences of time derivatives.}

Let us discuss some aspects of the classical physics of the theory
we are considering, and its possible regimes.  In particular we want
to point out the existence of a regime of the theory to which we will
make explicit reference in the last section.  To this aim, consider
the energy balance equation, namely the component of the equations
of motion (\ref{eqmot}) normal to the constant-$t$ surface (or the
scalar hamiltonian constraint of the non-gauge fixed theory). This
equation --which is equivalent to
equations (\ref{N}) or  (11)-- can be written as
\f
	\rho_{gr} =  \rho_{matter} - \rho_\lambda
\label{eb}
\ff
where: $\rho_{matter}$ is the energy density of the scalar field
\f
	\rho_{matter} = N^{-2} T_{00} = {N^{-2}\over 2} \partial_0\phi
\partial_0\phi = {N^{-2}\over 2} \mu^2;
\ff
 $\rho_\lambda$ is the absolute value of the (negative) cosmological
energy density
\f
	\rho_\lambda = {\lambda\over 16\pi G};
\ff
and $\rho_{gr}$ is the ``gravitational energy density'', namely the
ADM scalar constraint of pure gravity
\f
	\rho_{gr} = {16\pi G\over q} (p_{ab}p^{ab}-p^2)
	+ {R\over 16\pi G} = C_{ADM}.
\ff
In (\ref{eb}), $\rho_\lambda$, as well as  $\rho_{matter}$, are
non-negative.   Therefore, for suitable initial data, these two terms
may
cancel --or approximately cancel.
We denote such set of initial data
as the ``balanced clock regime --or, if they cancel approximately,
the
``approximate balanced clock regime''.   The cosmological term
induces Einstein's cosmic repulsion, while the energy
density of the ``clock'' field tends to crumple space-time. The two
can
be balanced.  In the balanced clock regime the gravitational
variables satisfy the constraint corresponding to pure gravity.  In a
sense (and for a short time interval), the gravitational field can be
unaware of the existence of the clock field and of the cosmological
constant: the two balance out.   Notice that this situation is
compatible with arbitrary initial conditions of the gravitational
field corresponding to a pure gravity field.  We will be particularly
interested in this regime because it is the regime in
which the theory we are studying better mimics pure gravity.  In the
``approximate balanced clock regime'' the cosmological term in the
Hamiltonian is, by definition,
much larger than the pure gravity term, which is
zero, or close to zero, and therefore
we can view the Hamiltonian (\ref{hc}), which can be written as
\f
	H = \sqrt{2}\mu\int d^3x\ \sqrt{q}\ \sqrt{{\lambda\over
16\pi G} - C_{ADM}},
\label{q}
\ff
as formed by the ``unperturbed term'' given by the cosmological term
plus a small perturbation given by $C_{ADM}$: the second term in the
square root is much smaller than the first.

In order to increase the readability of our equations,
we now choose a value for the constant $\mu$. A readjustment of
this
value determines only a rescaling of the coordinate time $t$, which
is of course arbitrary.  It is clear that if we could fix $\mu$ in such
a way that the coordinate time $t$ be precisely equal to the proper
time, then the legibility of our results be greatly improved. It
is not possible to achieve this result generically (because the
relation between the flow of proper time and the flow of $\phi$ is
determined dynamically and we cannot impose it), but it is possible
to choose $\mu$ in such a way that $t$ is the proper time in
particularly interesting physical regimes.  In fact, $t$ is the proper
time if the Lapse function is 1, or, using (\ref{N}) if
\f
	k_{ab} k^{ab}-k^2 + \mu^2  8\pi G = \lambda - R.
\ff
In the exact balanced clock regime this gives
\f
	\mu^2  = {\lambda\over 8\pi G},
\label{mu}
\ff
which is the value of $\mu$ that we shall definitely assume from
now on.  With this choice of the value of $\mu$, we have that in the
balanced clock regime the Hamiltonian evolves the system in proper
time.   Inserting this value of $\mu$ in (\ref{q}) we have the
Hamiltonian
\f
	H = 2\rho_\lambda \int d^3x\ \sqrt{q}\ \sqrt{1 - {C_{ADM}
\over
\rho_\lambda}}.
\ff
An observation that will play a major
role below is that the unperturbed term
\f
	H_0 = {\lambda\over 8\pi G} \int \sqrt{q} = 2\rho_\lambda V
\ff
is --up to a constant-- the total volume $V$.

Before going to the quantum theory, we shift to the Ashtekar
variables \cite{abhay}.  In terms of the Ashtekar variables, the
theory we are considered is defined as follows.  The phase space is
coordinatized by the Ashtekar conjugate variables $A^i_a$ and
$\tilde E^{ai}, \ i=1,2,3$, where $A^i_a$ is the space projection of
the selfdual spin connection and $\tilde E^{ai}$ is the densitized
controvariant triad, subjected to the first class Ashtekar's gauge
and diffeomorphism constraints  \cite{abhay}.  The Ashtekar's
hamiltonian constraint, which in the presence of the  cosmological
constant is
\f
	\tilde{\tilde{C}\,}_{Ashtekar}^{(\lambda)} =
{\epsilon_{ijk}\over 16\pi G}\ \left( F^i_{ab} \tilde E^{aj}
\tilde E^{bk} - {1\over 3!} \lambda \epsilon_{} \tilde E^{ai} \tilde
E^{bj} \tilde E^{ck}\right)
\ff
is absent, and the dynamical evolution is generated by the
Hamiltonian
\f
	H = \sqrt{2}\mu\int d^3x \ \sqrt{-
	\tilde{\tilde{C}\,}_{Ashtekar}^{(\lambda)} }.
\label{hamiltonian}
\ff

\vfill\eject

\section{Quantum Theory}

We now construct the quantum theory corresponding to General
Relativity in the gauge described above.  We refer to \cite{report},
for the definition of the classical loop variables $T[\alpha]$ and
$T^a[\alpha](s)$ in terms of the Ashtekar variables, and for the
definition of the corresponding quantum operators $\hat T[\alpha]$
and $\hat T^a[\alpha](s)$.  We refer to \cite{spinnet}, for the
definition of the spin network basis (that solves the theory's
Mandelstam identities) and the $s\,$-knot states $|s\rangle$ that
solve
the diffeomorphism constraint, and to
\cite{volume,hamiltonian,roumen}  for the construction of the
quantum operator corresponding to the hamiltonian
(\ref{hamiltonian}).  Here, we briefly review the results of these
references, in order to fix conventions and notation, and for
completeness. We start from
the conventional vector Ashtekar variables $A_a^i$ and $\tilde
E^{ai}$, as defined in \cite{abhay,abhaybook}. We denote the Pauli
matrices as $\sigma_i$.  The spinorial Ashtekar connection is
defined by
\f
		A_a = -{i\over 2} A_a^i \sigma_i
\ff
(Ashtekar convention). The spinorial triad is defined by
\f
		\tilde E^{a} = - 2 i \tilde E^{ai} \sigma_i
\ff
(Iwasaki convention). Given a loop $\alpha: S_1\rightarrow M$ with
components $\alpha: s\in[0,2\pi]\longmapsto\alpha^a(s)$, the
Ashtekar parallel propagator matrix $U_\alpha(t,s)$ along $\alpha$
is the path ordered exponential of the Ashtekar connection
along the loop, namely the $SL(2,C)$ matrix defined by
\begin{eqnarray}
		 {d\over ds}\  U_\alpha(t,s)
		&=&  U_\alpha(t,s)\ {d\alpha^a(s)\over ds}
A_a(\alpha(s)),
\\
		\lim_{s\rightarrow t+}U_\alpha(t,s) &= & 1.
\end{eqnarray}
We indicate $\lim_{s\rightarrow t-}U_\alpha(t,s)$ (namely the
parallel transport all around the loop) as $U_\alpha(t)$.  The loop
observables are defined by
\begin{eqnarray}
		T[\alpha] & =& Tr[U_\alpha(0)] \\
		T^a[\alpha](s) &=& Tr[U_\alpha(s) E^a(\alpha(s))] \\
		T^{ab}[\alpha](s,t) &=& Tr[U_\alpha(s,t)
			E^a(\alpha(t))U_\alpha(t,s) E^b(\alpha(s))]
\end{eqnarray}
with obvious generalizations for the loop observables with three or
more indices. (Notice the absence of the factor
${1\over 2}$ used
in some papers; the  ${1\over 2}$ factors are rather disturbing when
using spin network states.)\  The loop representation of the Poisson
algebra of
the loop variables is defined on a linear space of functionals
$\psi(\gamma)$ of multiple loops $\gamma$ (set of a finite number
of loops).  The algebraic dual of this space (the bras) is spanned by
the loop states $\langle\gamma|$, defined by
\f
		\langle\gamma|\psi\rangle=\psi(\gamma).
\ff
The representation is given as follows
\begin{eqnarray}
		\hat T[\alpha] \psi(\gamma) &=&\psi(\gamma\cup\alpha)
\\
		\hat T^a[\alpha](s) \psi(\gamma)  &=&
		\Delta^a[\gamma,\alpha(s)]
		\left(
			\psi(\gamma\#\alpha)-\psi(\gamma\#\alpha^{-1})
		\right)
\\
		\hat T^{ab}[\alpha](s,t)  \psi(\gamma)  &=&
		\Delta^a[\gamma,\alpha(s)] \Delta^b[\gamma,\alpha(t)]
		\  \sum_{i=1}^4 \psi(\alpha\#_{st}^i\gamma)
\label{operators}
\end{eqnarray}
where $\gamma\#\alpha$ is the loop obtained going around
$\gamma$ and then around $\alpha$, and the four loops
$\alpha\#_{st}^i\gamma$ are obtained rerouting the two
intersections (at $s$ and $t$) in all possible ways. For more details,
see \cite{report}.  The distributional factor is
\f
	 \Delta^a[\gamma,x]
	= l_p^2 \int ds\ {d\gamma^a(s)\over ds}\ \delta^3(\gamma(s),
x)
\ff
where $l_p=\sqrt{\hbar G}$ is the Planck length.  The loop basis is
overcomplete \cite{report}. A complete and non overcomplete basis
is the spin network basis, defined in \cite{spinnet}.   In this paper,
we will only consider the sector of the quantum theory defined by
the {\it trivalent\,} spin network states. The extension to higher
order spin networks implies additional algebraic complexity, which
has not been entirely worked out yet.
A trivalent spin network $S$ is an
oriented colored trivalent (namely with three links per node)
graph, imbedded into the space manifold, in
which the coloring (assignment of non-negative integers to each link
and each
node of the graph) satisfies certain conditions at each node:
the sum of the coloring of the three links incidents on each
node is even, and none of the three coloring is larger than the
sum of the other two \cite{penrose}.  These conditions are
equivalent
to the
requirement that it is possible to decompose the graph into
a family of
closed loops as follows: replace  each link colored $l$ with $l$
overlapping segments and pair-wise join the segments at each node,
in such a way that no two segments belonging to the same link are
joined.
Such a decomposition is not unique, since in general there are
many ways of joining segments, namely of routing the loops through
the nodes of the graph. If $\gamma_1, ..., \gamma_M$ are all the
possible decompositions of a spin network $S$, then the spin
network state $\langle S|$ is defined by
\f
	\langle S| = \sum_j (-1)^{n_j+c_j+1}\ \langle \gamma_j|
\label{sn}
\ff
where $n_j$ is the number of single loops in $\gamma_j$ and $c_j$
is the number of crossings in an arbitrary planar representation of
$\gamma_j$ (so that the sum produces in fact an
anti-symmetrization of the loops along each link; the overall sign is
determined by the orientation of the spin network
\cite{spinnet}).    It can be
shown that the spin network states (including higher order spin
networks) form a non-overcomplete basis \cite{spinnet}.

Since the spin network states form a basis, a ket state
$|\psi\rangle$ is completely characterized by the quantities
$\langle S|\psi\rangle$, which from now on we denote as $\psi(S)$.
Since the basis is not overcomplete, any assignment of quantities
$\psi(S)$ determines a state of the theory.  In particular, we may
define spin network characteristic (ket) states $|\psi_S>$ by
\f
		\psi_S(S') = \left\{\begin{array}{ll}
					= 1\ \ &{\rm if} \ S=S' \\
					= 0 \ \ &{\rm otherwise}.
					\end{array}
				\right.
\ff
The action of the elementary loop operators (\ref{operators}) on the
spin network states is directly computed from (\ref{operators}) and
(\ref{sn}).   The diffeomorphism constraint can then be solved easily.
The solution are labeled by the $s\,$-knots, namely by the
diffeomorphism equivalent classes of imbedded spin networks,
which we denote as $s$.  For each class $s$, there is a (ket)
state $\psi_s$ that solves the diffeomorphism constraints, defined
by
\f
		\psi_s(S) = \left\{\begin{array}{ll}
					= 1\ \ &{\rm if} \ S\in s \\
					= 0 \ \ &{\rm otherwise}.
					\end{array}
				\right.
\ff
We indicate the state $\psi_s$ also as $|s\rangle$.  These states are
labeled by knotted and linked (sets of) colored graphs.  Each of
these states represents an independent diffeomorphism invariant
physical quantum state of the gravity+scalar field theory.
Notice that the $s\,$-knots $s$, which label the physical
quantum states of
the gravitational-scalar field system
are not imbedded in space, they are
abstract objects in the same
sense in which knots of knot theory are.
 We recall
here that Roger Penrose introduced spin networks precisely in an
attempt to describe quantum geometry.  In the present
formalism, the abstract spin networks ($s\,$-knots) describe
precisely
the quantum states of the geometry.  The $s\,$-knot states carry
more
information than in Penrose's original version: first, they carry
information about their knotting and linking, second, they may have
intersections of order higher than three.  The key addition with
respect to Penrose's construction, of
course, is the knowledge of the set of quantum operators
acting on the space spanned by these quantum states.

We now come to the second main assumption that define the
T-theory.
We promote the state space we have defined to an Hilbert space by
choosing a scalar product.  This is uniquely determined by requiring
the $s\,$-knot states to be orthonormal
\f
			\langle s |s' \rangle = \delta_{s\,s'}\ .
\label{scalar}
\ff
We discuss this choice in the last section.  We just notice here that,
since the states $|s\rangle$ are linearly independent, the definition
(\ref{scalar}) is consistent, unlike earlier preliminary suggestions
to postulate that the knot states be orthonormal \cite{report}.

In order for an operator to be well defined on the space of the
diffeomorphism invariant physical quantum states, this
operator must be diffeomorphism invariant. The next step in the
definition of the theory is thus to recognize diffeomorphism
invariant observables, express them in terms of the loop operators,
and compute their action on the $s\,$-knot states.  Since
diffeomorphism invariant quantities are in general non-trivial (and
in particular non-linear) functions of the elementary fields, the
construction faces the difficulty of defining operator products.
Conventional regularization procedures fail, in general, in the
present context, because they almost invariably break
diffeomorphism invariance.  Diffeomorphism invariant
regularization techniques have then been
developed in \cite{review-ls,weaves,volume,hamiltonian}.  These
techniques can be viewed as diffeomorphism invariant analogs of
normal-ordering prescriptions. Using these techniques, various
operators of physical interest have been constructed.    Once the
regularization procedure has been chosen, the computation of the
action of the operators on the states is a tedious but
straightforward exercise.

In particular, the volume operator, which
corresponds to the classical observable
\f
		V = \int d^3x\ \sqrt{\det q}
\ff
has been constructed in \cite{volume}, and shown to be diagonal in
the spin network basis. It acts on the trivalent states as follows
\f
		\hat V \ |s \rangle = {1\over 4}\l_p^3\ \sum_{i\in s}
		\sqrt{\hat v_i}	\
		 |s \rangle;
\label{vol}
\ff
here the index $i$ labels the nodes of $s$ and the operator $\hat
v_i$ is defined by
\f
	\hat v_i\  |s \rangle = v_i \ |s \rangle
\ff
where
\f
	v_i= {a_ib_ic_i+a_ib_i+b_ic_i+c_ia_i}
\label{w}
\ff
The integers $a_i, b_i, c_i$ are defined by
\f
	p_i=a_i+b_i,\ \ \ \  q_i=b_i+c_i, \ \ \ \  r_i=c_i+a_i,
\label{abc}
\ff
where $p_i, q_i$ and $r_i$ are the colors of the three links adjacent
to the node $i$. Geometrically, $a_i$  is the number
of segments routed through $i$ between the $p_i$ and  the $r_i$
link, $b_i$  is the number
of segments routed between the links $p_i$ and $q_i$,
and so on.   Therefore, an $s\,$-knot state $|s\rangle$ is an
eigenstate of the volume with eigenvalue
\f
	 V(s) =  {1\over 4}\l_p^3\  \sum_{i\in s}
	\sqrt{a_ib_ic_i+a_ib_i+b_ic_i+c_ia_i}.
\ff

The Hamiltonian operator  $\hat H$ corresponding to the Hamiltonian
(\ref{hamiltonian}) has been constructed in
\cite{hamiltonian}.  Its action on the trivalent states is given by
\begin{eqnarray}
	\hat H  |s \rangle &=& \sqrt{{2\mu^2\over 16\pi G}} \
	\sum_i \sqrt{\l_p^4 Z\hat M_i + {\lambda l_p^6\over 16}\hat
	v_i}
	\ |s \rangle
\nonumber \\
	&=&\rho_\lambda l_p^3 \sum_i
	\sqrt{\hat v_i+\alpha \hat M_i}\ |s \rangle
\label{ham}
\\
	\alpha &=& {16Z\over\lambda l_p^2}
\label{peppo}
\end{eqnarray}
$Z$ is an arbitrary (finite) renormalization constant. Notice that
$\alpha$ is dimensionless. The action of the operator $\hat v_i$ on a
node
$i$ has been given above.  The hamiltonian node operator $\hat M_i $
acts on a trivalent node as follows
\f
	\hat M_i = \sum_{l=1,2,3}\ \sum_{\epsilon=1,-1} \
	\sum_{\epsilon'=1,-1}
	A_{\epsilon\epsilon'}(p_l,q_l,r_l)
	 \ \hat D_{i;l\epsilon\epsilon'};
\label{lee}
\ff
the index $l$ labels the three links adjacent to the node $i$;
$r_l$, $p_l$ and $q_l$ are the colors of these three links,
in the following order: $r_l$ is the color of the link $l$;
$p_l$ the next one and $q_l$ the last one, in the order given by the
orientation of the spin network.  Finally, $\hat
D_{i;l\epsilon\epsilon'}$ is the operator that acts on an $s\,$-knot
node
by: (i) creating two additional nodes, one along each of the two links
-different from $l$- adjacent to the node $i$, (ii) creating a novel
link, colored 1, joining these two nodes, (iii) assigning the coloring
$p_l+\epsilon$ and, respectively, $q_l+\epsilon'$ to the links that
join the new formed nodes with the node $i$. This is illustrated in
Figure 1.
\vskip.5cm
\begin{picture}(387,114)(105,606)
\thicklines
\put(211,654){\line(-5,-3){ 60}}
\put(211,654){\line( 5,-3){ 60}}
\put(211,654){\line( 0, 1){ 60}}
\put(172,618){\makebox(0,0)[lb]{\raisebox{0pt}[0pt][0pt]{\twlbf r}}}
\put(201,694){\makebox(0,0)[lb]{\raisebox{0pt}[0pt][0pt]{\twlbf q}}}
\put(244,618){\makebox(0,0)[lb]{\raisebox{0pt}[0pt][0pt]{\twlbf p}}}
\put(105,654){\makebox(0,0)[lb]{\raisebox{0pt}[0pt][0pt]{\svtnbf
\^D}}}
\put(120,651){\makebox(0,0)[lb]{\raisebox{0pt}[0pt][0pt]{\twlbf
r$+-$}}}
\put(431,685){\line( 2,-3){ 40.462}}
\put(431,648){\line(-5,-3){ 60}}
\put(431,648){\line( 5,-3){ 60.735}}
\put(431,648){\line( 0, 1){ 60}}
\put(311,654){\makebox(0,0)[lb]{\raisebox{0pt}[0pt][0pt]{\svtnbf
=}}}
\put(386,609){\makebox(0,0)[lb]{\raisebox{0pt}[0pt][0pt]{\twlbf r}}}
\put(482,606){\makebox(0,0)[lb]{\raisebox{0pt}[0pt][0pt]{\twlbf p}}}
\put(419,703){\makebox(0,0)[lb]{\raisebox{0pt}[0pt][0pt]{\twlbf q}}}
\put(453,657){\makebox(0,0)[lb]{\raisebox{0pt}[0pt][0pt]{\twlbf 1}}}
\put(403,657){\makebox(0,0)[lb]{\raisebox{0pt}[0pt][0pt]{\twlbf
q$-$1}}}
\put(432,621){\makebox(0,0)[lb]{\raisebox{0pt}[0pt][0pt]{\twlbf
p$+$1}}}
\end{picture}

\vskip.3cm
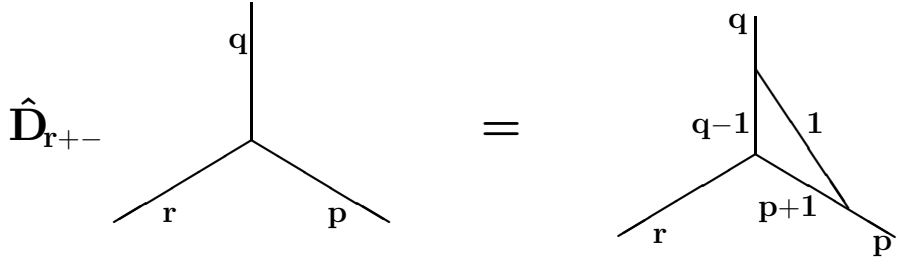
\centerline{Figure 1: Action of $\hat D_{i;l\epsilon\epsilon'}$.}
\vskip1cm

 The explicit computation of the $A_{\epsilon\epsilon'}$
coefficients has been recently concluded \cite{roumen}, giving
\begin{eqnarray}
		A_{\epsilon\epsilon'}(p,q,r)
		&=& {B_{\epsilon\epsilon'}(p,q,r)\over (p+1)(q+1)};
\nonumber \\
		B_{++}(p,q,r) &=& pq,    \nonumber \\
		B_{+-}(p,q,r)  &=& -(q+r)pc,   \nonumber  \\
		B_{-+}(p,q,r)  &=& -(p+r)qa,    \nonumber   \\
		B_{--}(p,q,r)  &=& (p+2)(q+2)(pq+b-ac),
\label{A}
\end{eqnarray}
where $a,b$ and $c$ are defined as in ({\ref{abc}}).

While the  general structure and the coefficients of the action of the
operator  $\hat M_i$ are determined by the classical Hamiltonian
and by the  regularization procedure, there is nevertheless freedom
in  defining the precise geometrical action of the $\hat D$ operator.
This freedom is a conventional quantum mechanical ordering
ambiguity.  Some alternatives are ruled out immediately because
they yield trivial or inconsistent hamiltonian operators; to some
extent the remaining ambiguity is resolved by the requirement of
preserving gauge invariance: the geometrical action must be well
defined on knot classes.  While no credible alternative ordering is
known at the moment, there  is no reason to believe that the one
considered here is unique.  Therefore, the choice of the ordering
given by  $\hat D$ represents  another assumption in the process of
constructing the  diffeomorphism invariant T-theory.

There is a remaining open problem before having the explicit action
of the hamiltonian operator on an arbitrary trivalent $s\,$-knot: to
compute the operator square root in (\ref{ham}).  The next section is
devoted to solve this problem.

\section{Square root}

This section is rather technical. It derives the missing technical
ingredient for the computation of transition amplitudes in the
T-theory, namely the extraction of the square root in
the  Hamiltonian
operator.  To this aim, let us introduce an index $\mu=1...12$ as a
collective index for the indices $l$, $\epsilon$ and $\epsilon'$ in
(\ref{lee}).  We also use $A^\mu(i)= A_{\epsilon\epsilon'}
(p_l,q_l,r_l) $.  We indicate a node $i$ as $|i\rangle$ and
\f
	|i,\mu\rangle=\hat D_{i,\mu}\ |i\rangle
\ff
indicates the portion of the $s\,$-knot that replaces $|i\rangle$
after
the action of the operator.   Adopting Einstein's convention on
repeated $\mu-$indices, we then can write (\ref{lee}) as
\f
		M_i\ | i\rangle=A^\mu(i)\  |i, \mu\rangle.
\ff
Our task is to extract the square root in (\ref{ham}). We
seek an operator $\hat H_i$ such that
\f
	\hat H_i = \sqrt{\hat v_i + \alpha \hat M_i },
\label{r}
\ff
namely, such that
\f
	\hat H_i^2 |i\rangle = v_i |i\rangle + \alpha
	A^\mu(i)|i,\mu\rangle,
\label{uno}
\ff
where the number $v_i$ is given in (\ref{w}). The key observation is
that $\hat H_i^2$ is --in a sense-- lower triangular, and therefore
we
can
seek for $\hat H_i$ of this same form.  We thus assume
the following form for $\hat H_i$
\f
	\hat H_i |i\rangle = a(i) |i\rangle + \alpha a^\mu(i)
|i,\mu\rangle
	+\alpha^2 a^{\mu\nu}(i) |i,\mu\nu\rangle + ....
\label{due}
\ff
where $|i,\mu\nu\rangle$ is given by
\f
	|i,\mu\nu\rangle =\hat D_{i,\nu} \hat D_{i,\mu}\ |i\rangle,
\ff
and so on. We can further simplify the notation by dropping the index
$i$ everywhere, since everything here is happening around a fixed
intersection $i$.    We now compute the coefficients $a=a_i$,
$a^\mu=a^\mu(i)$, $a^{\mu\nu}=a^{\mu\nu}(i)$, ... by comparing
(\ref{uno}) and (\ref{due}).  By acting twice on the state $|\rangle=
|i\rangle$ with $\hat H=\hat H_i$, we obtain
\begin{eqnarray}
	\hat H \hat H |\rangle &=& \hat H
	 \left[a |\rangle + \alpha a^\mu |\mu\rangle
	+ \alpha^2 a^{\mu\nu} |\mu\nu\rangle + ...\right]
\nonumber \\ &=&
	 a \left(a |\rangle + \alpha a^\mu |\mu\rangle
	+ \alpha^2 a^{\mu\nu} |\mu\nu\rangle + ... \right)
\nonumber \\ &&
	+ \alpha a^\mu \left(a(\mu) |\mu\rangle + \alpha a^\nu(\mu)
|\mu\nu\rangle
	 + ....\right)
\nonumber \\ &&
	+ \alpha^2 a^{\mu\nu}(\mu) v(\mu\nu) |\mu\nu\rangle
\nonumber \\ &&  + ...
\nonumber \\ &=&
	a^2 |\rangle + \left(a +a(\mu)\right) \alpha a^\mu|\mu\rangle
\nonumber \\ && +
	\left[\left(a+a(\mu\nu)\right) \alpha^2
	a^{\mu\nu}+\alpha^2 a^\nu(\mu)a^\mu\right] |\mu\nu\rangle +
...
\end{eqnarray}
Where $v(\mu)$ indicates the value of $v_i$ on the node $i$
after the action of $\hat D_{i,\mu}$ (the coloring of two of the links
has been altered by a unit:
$\epsilon$ or $\epsilon'$); similarly, $a^\nu(\mu)$ is the
coefficient $a^\nu$ acting on the same altered node.  We now
equate state by state with (\ref{uno}); notice that these
equalities determine a sufficient but not necessary condition for
(\ref{r}) to hold, because the $ |\mu\nu...\rangle$'s are not linearly
independent in general.  For the moment, however, we are not
concerned with uniqueness. We obtain
\begin{eqnarray}
	a &=& \sqrt{v} 			\label{a} \\
	a^\mu &=&{A^\mu\over a+a(\mu)}    \label{b}\\
	a^{\mu\nu}& =&{a^\mu a^\nu(\mu)\over
	a+a(\mu\nu)}.
\end{eqnarray}
By repeating this procedure for higher orders, it is easy to conclude
that the general form of the coefficients is
\begin{eqnarray}
	& a^{\mu_1 ... \mu_n} = &
\nonumber \\ &
 {a^\mu_1
a^{\mu_2...\mu_n}(\mu_1)+
	a^{\mu_1\mu_2} a^{\mu_3...\mu_n}(\mu_1\mu_2)+ ... +
	a^{\mu_1...\mu_{n-1}}a^\mu_n(\mu_1...\mu_{n-1})\over
	a+a(\mu_1...\mu_n)}&
\label{c}
\end{eqnarray}
Equations (\ref{a},\ref{b},\ref{c}) express all
the coefficients $a^{\mu_1...\mu_n}(i)$ of the Hamiltonian, in terms
of the quantities $v(i)$ and $A^\mu(i)$.  The node Hamiltonian is
then
\f
	\hat H_i = \sum_{n=0}^\infty \alpha^n\ a^{\mu_1...\mu_n}(i)
	\ |i, \mu_1...\mu_n\rangle
\label{exp}
\ff
And the full Hamiltonian is
\f
	\hat H |s\rangle = {\lambda\over 8\pi G}
	\ l_p^3\  \sum_i \hat H_i\ |s\rangle
\label{HH}
\ff

Since the evolution generated by the classical Hamiltonian breaks
down at finite times, we do not expect the quantum Hamiltonian
operator to be finite and well defined on all states.
On which states is it ill defined ?
There are two potential sources of difficulties: one is of course the
infinite sum in the node Hamiltonian
(\ref{exp}), which may not converge on certain
states. But there is another one which we should discuss. In the
definition  (\ref{c}) of the coefficients $a^{\mu_1...\mu_n}(i)$ we
divide by the (square root of the) volume eigenvalues $v(i)$ of the
intersection $i$.  Looking at equation (\ref{w}), we see that there is
a degenerate case in which the $v(i)$ may vanish. This is the case in
which two out of three numbers $a, b, c$ vanish (they cannot all
vanish). In turn, this case corresponds to a (degenerate) trivalent
intersection in which one of the three links has color zero, which is
to say a ``bivalent'' intersection, or a point of non differentiability
along the loop. These points were denoted ``kinks'' in early works on
the loop representation, and cannot be discarded a-priori from the
state space, because states with kinks are in the image of
operators such as the hamiltonian itself.  The operator (\ref{HH})
is ill defined on spin networks that contain such kinks. It is natural
to suspect that states with kinks correspond to points were the
clock regime breaks down, but such a conclusion is far from obvious,
and more insight is needed.  In particular, the ansatz that $H_i$ is
lower diagonal could simply be wrong when kinks are present.

Finally, we recall that ${\lambda\over 16\pi G}={1\over 2}\mu^2$ is
the energy density $\rho_{matter}$ of the clock field; therefore if
we denote the energy of the clock field per Planck volume as
$E_0={1\over 2}\mu^2 l_p^3$, we may write
\f
\hat H  = 2E_0 \sum_i \  \sum_{n=0}^\infty \
\alpha^n\, a^{\mu_1...\mu_n}(i) \ \hat D_{i;\mu_1}...\hat D_{i;\mu_n}.
\label{h}
\ff
The sum in $i$ is over all the nodes of the spin network $s$. The
coefficients $a^{\mu_1...\mu_n}(i)$ are explicitly given in
equations (\ref{w},\ref{A},\ref{a},\ref{b},\ref{c}), and are
dimensionless functions of the colorings. The model is defined by
two constants: $E_0$, which represents the energy
of the clock per Planck volume, and the dimensionless constant
$\alpha$, related to the ratio of the cosmological constant and
the Planck energy.
This completes the definition of quantum general
relativity in the clock gauge.

\section{Perturbation theory}\label{expansion}

By separating the first term from all the others in (\ref{h}), we can
rewrite the Hamiltonian as
\f
	\hat H  = {\lambda\over 8\pi G} \hat V +  \hat W,
\ff
where $\hat V$ is the volume operator and $\hat W$ is
\f
\hat W  = 2E_0 \sum_i \  \sum_{n=1}^\infty \
\alpha^n\, a^{\mu_1...\mu_n}(i) \ \hat D_{i;\mu_1}...\hat D_{i;\mu_n}.
\label{sum}
\ff
In the balanced clock regime discussed at the end of section 2, the
volume term dominates over the $\hat W$ term.  We can therefore
consider the possibility of viewing the term $\hat W$ as a
perturbation.
Notice that we do not need $\alpha$ to be small.
This is a particularly interesting program for
various reasons: first,  the balanced clock regime is the regime in
which our theory better approximates pure General Relativity;
second, the
eigenstates of the unperturbed Hamiltonian, namely the volume, are
completely known. Indeed, they are precisely the spin network
states that form the basis in which we are working.  Developing a
perturbation scheme is then straightforward. The idea of developing
a perturbation scheme around the volume as unperturbed Hamiltonian
was first suggested by L. Smolin \cite{ls}. In the unperturbed
theory, the time evolution of the basis states is given by
\f
	|s,t\rangle = e^{-i/\hbar E_s t} |s\rangle
\ff
where the energy  $E_s$ of the $s\,$-knot state is (from
(\ref{vol}) and (\ref{w}))
\f
	E_s = {E_0\over 2} \sum_{i\in s}
	\sqrt{a_ib_ic_i+a_ib_i+b_ic_i+c_ia_i},
\ff
and $t$ represents, thanks to the choice (\ref{mu}) of $\mu$, the
proper time. We can define an interaction picture by expanding a
generic time dependent state $|\psi,t\rangle$ in this time dependent
$s\,$-knot basis
\f
		|\psi,t\rangle = \sum_s \psi(s,t)\
		e^{-i/\hbar E_s t}\  |s\rangle.
\ff
Using conventional perturbation theory techniques, the
amplitude of a transition from a state $|s_i\rangle$ to a (different)
state $|s_f\rangle$ in a time $t$ is given by
\begin{eqnarray}
	\langle s_f,t|s_i,0\rangle &=& {-i\over\hbar}\int_0^t dt'
W_{fi}\ e^{i/\hbar(E_f-E_i)t}
\nonumber \\ &&
	+ \left({-i\over\hbar}\right)^2
	\int_0^t \int_0^{t'} dt' dt''
\nonumber \\ && \ \ \ \times \sum_n  W_{fn}W_{ni} \
	e^{i/\hbar(E_f-E_n)t'}\, e^{i/\hbar(E_n-E_i)t''}
\nonumber \\ &&
 + ...
\label{pippo}
\end{eqnarray}
where we introduced the notation
\f
	W_{ab} = \langle s_a|W|s_b\rangle
\ff
for the matrix elements of the perturbation term.  Since these are
time independent, we can perform the integrations, giving
\begin{eqnarray}
	\langle s_f,t|s_i,0 \rangle &=&
	-W_{fi} {e^{i/\hbar(E_f-E_i)t}-1\over  E_f-E_i}
+	\sum_nW_{fn}W_{ni}
\nonumber \\ &&
	\times \left[{e^{i/\hbar(E_f-E_i)t}-1\over
(E_n-E_i)(E_f-E_i)} -{e^{i/\hbar(E_f-E_n)t}-1\over (E_n-E_i)(E_f-
E_n)}\right]
\nonumber \\ &&
+ ...
\end{eqnarray}
Up to now, we have made no approximations.   For small time, we can
expand the exponentials, giving
\f
	\langle s_f,t|s_i,0\rangle =
	-W_{fi} {i\over\hbar} t+O(t^2)
\ff
So that, to first order in $t$, the transition probability is
\f
	P_{i\rightarrow f} = |\langle s_f,t|s_i,0\rangle |^2 =
	{1\over\hbar^2} |W_{fi}|^2 t^2
\ff

If we further restrict our conditions and assume that $\alpha$ is
small, then the first term in the sum (\ref{sum}) dominates over the
others,
and we can write
\begin{eqnarray}
	W_{fi} &=& \langle s_a|2 E_0 \alpha \sum_i a^\mu(i)\hat
	D_{i;\mu} |s_b\rangle    \ +O(\alpha^2)
\nonumber \\
	&=& \langle s_a|2 E_0 \alpha \sum_i
{A^\mu(i)\over a(i)+a(i,\mu)}
\hat 	D_{i;\mu} |s_b\rangle  \ +O(\alpha^2)
\end{eqnarray}
The physical meaning of taking $\alpha$ small is that the
cosmological
constant energy density and the clock energy density (which, under
our
assumptions balance each other) are both large compared with the
Planck energy density.  This is perhaps a rather unrealistic
assumption, and
we consider it here only for illustrative purposes.  As an example,
let us
compute a simple transition probability in the approximations
considered.
Let $s_i$ be the $s\,$-knot formed by two nodes, connected
by three links,
with
colors 1,1 and 2.  Let $s_f$ be the $s\,$-knot formed by 4 nodes
(arranged as the
vertices of a tetrahedron), with the couples of opposite (non
adjacent) links having colors (1,1), (2,2) and (3,1).  See Figure 2.
\vskip.7cm
\setlength{\unitlength}{0.009in}
\begin{picture}(474,185)(165,460)
\thicklines
\put(219,592){\oval(108,106)}
\put(219,645){\line( 0,-1){107}}
\put(276,592){\makebox(0,0)[lb]{\raisebox{0pt}[0pt][0pt]{\tenbf 1}}}
\put(168,592){\makebox(0,0)[lb]{\raisebox{0pt}[0pt][0pt]{\tenbf 1}}}
\put(222,592){\makebox(0,0)[lb]{\raisebox{0pt}[0pt][0pt]{\tenbf 2}}}
\put(212,512){\makebox(0,0)[lb]{\raisebox{0pt}[0pt][0pt]{\tenbf s}}}
\put(219,504){\makebox(0,0)[lb]{\raisebox{0pt}[0pt][0pt]{\ninbf i}}}
\put(397,592){\oval(108,106)}
\put(397,645){\line( 0,-1){107}}
\put(397,592){\line( 1, 0){ 53}}
\put(346,592){\makebox(0,0)[lb]{\raisebox{0pt}[0pt][0pt]{\tenbf 1}}}
\put(433,545){\makebox(0,0)[lb]{\raisebox{0pt}[0pt][0pt]{\tenbf 1}}}
\put(400,558){\makebox(0,0)[lb]{\raisebox{0pt}[0pt][0pt]{\tenbf 2}}}
\put(417,595){\makebox(0,0)[lb]{\raisebox{0pt}[0pt][0pt]{\tenbf 1}}}
\put(400,612){\makebox(0,0)[lb]{\raisebox{0pt}[0pt][0pt]{\tenbf 3}}}
\put(437,618){\makebox(0,0)[lb]{\raisebox{0pt}[0pt][0pt]{\tenbf 2}}}
\put(390,512){\makebox(0,0)[lb]{\raisebox{0pt}[0pt][0pt]{\tenbf s}}}
\put(397,504){\makebox(0,0)[lb]{\raisebox{0pt}[0pt][0pt]{\ninbf f}}}
\put(584,592){\oval(108,106)}
\put(584,645){\line( 0,-1){107}}
\put(584,592){\line( 1, 0){ 55}}
\put(534,592){\makebox(0,0)[lb]{\raisebox{0pt}[0pt][0pt]{\tenbf 1}}}
\put(620,545){\makebox(0,0)[lb]{\raisebox{0pt}[0pt][0pt]{\tenbf 1}}}
\put(588,558){\makebox(0,0)[lb]{\raisebox{0pt}[0pt][0pt]{\tenbf 2}}}
\put(605,595){\makebox(0,0)[lb]{\raisebox{0pt}[0pt][0pt]{\tenbf 1}}}
\put(578,512){\makebox(0,0)[lb]{\raisebox{0pt}[0pt][0pt]{\tenbf s}}}
\put(620,619){\makebox(0,0)[lb]{\raisebox{0pt}[0pt][0pt]{\tenbf 2}}}
\put(586,610){\makebox(0,0)[lb]{\raisebox{0pt}[0pt][0pt]{\tenbf 1}}}
\put(584,508){\makebox(0,0)[lb]{\raisebox{0pt}[0pt][0pt]{\ninbf g}}}
\end{picture}

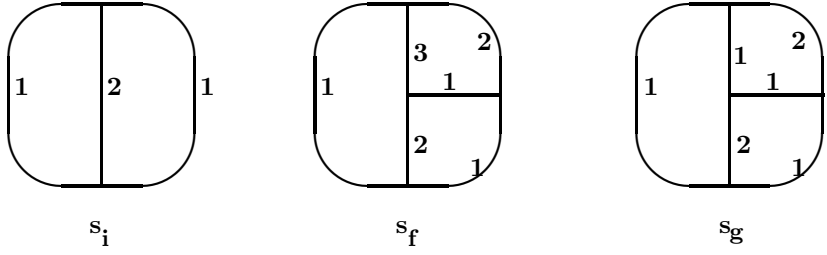
\centerline{Figure 2: Initial and final states}
\vskip.6cm

We begin by computing the action of the first term of the
perturbation series
on $|s_i\rangle$. The
sum over the nodes gives only a factor of 2, for symmetry.  On each
node, there are three terms produced (one per couple of links). Two
of these are equal, again by symmetry, and it is easy to see that the
third fails to produce anything proportional to $s_f$. We thus have 4
equal terms (2 nodes -- 2 non-vanishing terms each).
Of the four terms ($++,+-,-+,--$) produced by each of
these 4, it is the ($++$) that yields a result proportional to $s_f$.
We have  in this case
\f
	r=1, \ \ \ p=2,\ \ \  q=1,
\ff
which yields
\f
	a=1,\ \ \ b=1,\ \ \ c=0.
\ff
from which we have
\f
	v(i)=\sqrt{abc+ab+bc+ca}= 1,
\ff
and
\f
	A_{++}={(p+2)(q+2)\over(p+1)(q+1)}(pq+b^2-ac)=6
\ff
After the action of the operator, the node has colors
\f
	r=1, \ \ \ p=3,\ \ \  q=3,
\ff
which yields
\f
	a=1,\ \ \ b=2,\ \ \ c=0.
\ff
from which we have
\f
	v(i\mu)=\sqrt{abc+ab+bc+ca}=\sqrt{2},
\ff
Bringing everything together we have finally
\f
	P_{i\rightarrow f} = {1\over\hbar}\ 2E_0\ \alpha\ 4\ {6\over
1+\sqrt{2}}\   t^2 =Z^2 {2^{10}9 \over (3+2\sqrt{2}) \pi^2}\
 {c^2\over l_p^2}\ t^2,
\ff
where we have reinserted the velocity of light $c\neq 0$ for clarity.
A completely analogous calculation gives
\f
	P_{i\rightarrow g} = Z^2 {2^{10}9 \over (3+2\sqrt{2}) 12^2
\pi^2}\  {c^2\over l_p^2}\ t^2.
\ff
for the transition probability to the state $|s_g\rangle$ with the
same graph as $|s_f\rangle$, but with the couples of opposite (non
adjacent) links having colors (1,1), (2,2) and (1,1). See Figure 2.
Therefore the
relative probability for $|s_i\rangle$ to make a transition (in a
short time) to $|s_f\rangle$ or to $|s_g\rangle$ is
\f
	{P_{i\rightarrow g} \over P_{i\rightarrow f} } = {1\over 144}.
\ff
What is interesting here, of course, are not the computed values, but
the fact that the above machinery allows us to compute qualitative
predictions of decay ratios.

\section{Discussion}

Using non-perturbative techniques, and in particular the
loop representation, we have a
constructed a quantum theory describing two interacting fields: the
gravitational field and a massless scalar field.\footnote{As made
clear by the hamiltonian analysis, the theory has three degrees of
freedom per space point: two gravitational and the scalar field's one.
In spite of its apparent disappearance in the gauge considered,
therefore, the physical scalar field is alive and well, and its
dynamics is fully taken into account. Despite immediate
appearances,
the quantum theory describes a quantum gravitational field {\it
and\,} a {\it quantum\,} scalar field.}  The main ansatzs on which the
theory  relies are the following.

\begin{itemize}

\item Scalar product: The weakest assumption of the theory is the
choice  (\ref{scalar}) of the scalar product.  The scalar product
chosen satisfies some  crucial  requirements: the operator
corresponding to the physical volume, for  instance,  is self-adjoint
with respect to this scalar product. On the other  hand, it is far
from clear, and perhaps dubious, that the scalar product
(\ref{scalar}) could satisfy all the necessary conditions for the
correct recovery of the classical limit. In particular, we recall that
the choice of the scalar product corresponds to the implementation
of the reality conditions in the quantum theory. The reality
conditions on the connection are non-trivial in the Ashtekar
formalism, and we do not know whether they are implemented  by
(\ref{scalar}).  The choice made may  very well  yield the Euclidean,
rather than Lorentian, General Relativity in the classical  limit
-- or none of the two.    There is a strict relation between the
problem of choice of the scalar product and possible
self-adjointness requirements on the Hamiltonian discussed below;
Smolin has suggested \cite{ls}
that the scalar product (40) could be modified
in order to make the Hamiltonian self-adjoint, and that this
may be done order by order in the perturbation expansion (the
zeroth order is already self-adjoint).
The effect of the non-triviality of the Ashtekar reality  conditions
on  the choice of the scalar product is being investigated by
Ashtekar  Lewandovski and  collaborators \cite{abhayetal}, and we
expect this investigation  to shed  light on this issue.  Here
we leave the problem open.  On the other side, notice that the doubts
about the correctness of the choice  (\ref{scalar}) do not bear on the
issue of the consistency of the  diffeomorphism invariant quantum
theory we are constructing; what  is at stake here is not the
consistency of the quantum theory, but  rather the possibility of
recovering the correct classical limit.

\item  Evolution:  The idea that we can describe evolution in a
covariant  fashion by evolving with respect to an arbitrary variable
is very old \cite{tradition}.  A potential difficulty with its
implementation in the  quantum context is that with
the ordering chosen, the Hamiltonian
fails to be a densely defined self-adjoint operator.  (This follows
from the observation that the Hamiltonian may raise, but may not
lower, the number of nodes, and therefore cannot be symmetric in
the -orthogonal- spin network basis. With a different ordering,
this is not necessarily the case. See below.).
Self-adjointness of the  Hamiltonian is not
necessary for the consistency of the conventional probabilistic
interpretation: non-unitary evolution is routinely employed in
quantum  mechanics to describe the dynamics of objects, for
example, having  a finite  probability of decaying.  In those cases the
failure of the evolution  to be  unitary simply signals the probability
that, say, the position of a  particle  may have no value at all at
some later time, because the particle has  decayed. In the present
case, the lack of unitarity of the evolution in  the  (rather arbitrary)
independent variable signals the probability that the  system will
not reach a later value of such a variable: the scalar field  may stop
increasing.  This has been discussed in detail elsewhere
\cite{carlotime}.  Whether or not such a procedure is viable is
a much debated issue \cite{christime}; we want to suggest that a
convincing solution of this issue
might come from exploring the consequences of various
proposed solutions within
a reasonably realistic
infinite dimensional theory as the one proposed here.
A separate problem, we believe, is
how the physical ``flowing'' time emerges from  a
generally  covariant quantum field theory, in which no
preferred time variable exists --the scalar field is an
arbitrarily chosen independent variable, not at all necessarily
connected with the ``perceived flow'' of physical time.
We are convinced that
this issue should be addressed in a
different context \cite{termo}, and is not relevant here.

\item Ordering: Alternative orderings of the Hamiltonian have to be
explored.  The simplest alternative is to consider the
symmetric part of the operator $\hat H$, which is a possible
step towards self-adjointness. Taking the symmetric part of
the Hamiltonian shouldn't affect the classical limit of the
theory, since the classical hamiltonian is real on the physical
solutions.  It would also be interesting to find  uniqueness
results, under the requirement of diffeomorphism  invariance of the
regularized operator. On the hamiltonian constraint see  \cite{hamc}.
Similarly, it may be that alternative definitions of the square root
are available.

\item Diffeomorphism invariance.  One should distinguish two
different kinds of invariance requirements.  On the one hand, the
theory should be gauge invariant in the sense that physical
quantities that are invariant under four-dimensional
diffeomorphisms must be identified and the theory should yield
expectation values for those quantities. The transition amplitudes
computed here satisfy this requirement.  In this sense, four
dimensional diffeomorphism invariance is implemented in the theory
in spite of the fact that the gauge has been fixed.  This is like saying
that QED predictions are gauge invariant, even if computed in the
Lorentz gauge.   On the other hand, the quantities computed here are
transition amplitudes with respect to
a specific evolution parameter, the scalar
field, arbitrarily chosen. This fact raises the question of the
relation between those quantities and quantities that represent
evolution with respect to a different evolution parameter.  This is a
very interesting issue, but we are not going to address it here.

\item Use of loop observables as fundamental observables:
Experience in  quantum field theory indicates that field operators
are too singular  to be  integrated in just one dimension, and one may
suspect that loop  operators cannot be defined in an interacting
quantum field theory.  This is a  strong objection against the loop
representation, but we believe that  this  objection overlooks an
aspect the theory: The loop operators are {\it  not}  defined in the
quantum theory.  They play only an intermediate role  at the
unconstrained level of the non-diffeomorphism invariant state
space.  Any  physical operator --as the hamiltonian operator
considered in this  paper-- is  integrated in three dimensions at
least.   Notice that the physical  states, namely the $s\,$-knot
states,  do not have support on loops,
but rather are formed by  diffeomorphism invariant
``extensions'' of loop states,
and can loosely be  thought as  a smearing over
all space --or, better over all smooth deformations-- of loop
states supported in one dimension.

\end{itemize}

The diffeomorphism invariance of general relativity  expresses  the
discovery that spatio-temporal location (locality) is physically
meaningful only in reference to other dynamical components of the
theory \cite{peter,whatis,tradition}.
The incorporation of  this aspect of
General Relativity into quantum field theory  requires a major step
ahead with respect to local quantum field  theory.  The theory
constructed in this paper is a crude attempt to take  this step  and
to sketch such a general relativistic quantum field theory.  The
distance  covered from local quantum physics is substantial:
quantum states,  operators  and the very notion of evolution appear
here in de-spatialized form:  spatial,  as well as temporal, location
are only defined relationally in the  T-theory.    We emphasize in
particular the unusual structure of the  quantum states. They have
no space-time  (nor momentum space) dependence. Rather, a state is
characterized  by topological and combinatorial relations only.  This
feature derives from the strict
implementation of three dimensional diffeomorphism invariance in
the quantum field  theory.  Because of diffeomorphism invariance
any genuine gauge  invariant  property   expresses solely relative
positions of physical structures  (among these, the gravitational
field); the combinatorial-topological structure of the $s\,$-knot
quantum  states expresses gauge invariant relative  positions in a
location-independent way.  This is complemented by  our treatment
of time, in which  temporal-location and evolution are replaced by
temporal location  and evolution relative to a physical field.  In
these two ways, the  quantum theory we are constructing is defined
in a general relativistic manner. Most of the previous attempts  to
combine general  covariance and quantum field theory (see for instance the
various contributions  in \cite{baez}, and in \cite{jmp})  have so far
been carried on only  in the restricted domain of theories with a
finite number of degrees  of  freedom --topological quantum  field
theories   \cite{topo}
are the most interesting of these attempts. (This has
produced
the curious and obviously wrong  popular belief that any fully
diffeomorphism invariant field  theory has a finite number
of degrees of freedom.)

The extent to which the T-theory is successful in this attempt is far
from clear to us.  One may consider three criteria of
evaluation of the theory: consistency, completeness, and  classical
limit. None of the three is clearly satisfied by the T-theory.
Consider  consistency.  State space and hamiltonian operator of the
T-theory are  well defined (up to the
incompleteness related to non-trivalent intersections).
We have shown that it is not difficult to
compute  quantum  transitions amplitudes to first order in some
approximation.  The  potential  problem is given by the divergences
that can appear at higher orders.   We do not know whether
divergences appear, and, in the  likely  case in which they do,
whether they can be controlled.  We consider  this as  the most
urgent problem to be investigated.    Next, consider the classical
limit issue.  The procedure of starting  from a  classical theory and
promoting it to a quantum theory by means of  some  quantization
prescription is designed to yield a quantum theory with  the  desired
classical limit.  However, it is not clear whether in our
construction  we have followed all steps of a complete quantization
prescription,  and there is at least one key check that we failed
to perform,  which is the consistency between the Hilbert structure
chosen  and the full set of classical reality conditions.
We just remark, once more, that even a --non-trivial--
theory with  the  incorrect classical limit may have interest as
an example of general relativistic quantum field theory.
Finally, as far as  completeness is  concerned, we recall that all
realistic physical quantum field theories  presently utilized  are
badly incomplete: We know such theories either via a
perturbation  expansion  that certainly does not cover the entire
relevant physics, or by means  of
non-perturbative
approaches, all of which provide  approximations to this or that
physical regime.   Therefore, we do not expect completeness in the
present context. The problem, is not so much completeness,  but
rather in which regimes (if any) is the theory predictive:  the  hope
expressed here is that the T-theory might represent a tool for
obtaining  physical predictions at the Planck scale.

We conclude with a list of problems that we think deserve further
investigation: (i) Study self-adjointness properties of other physical
observables, in order to check the scalar product (40) and study
possible modifications of (40). (ii) Study alternative orderings for
the Hamiltonian; search for a uniqueness result. (iii) Are there
alternative ways of defining the square root? (iv) What is the
physical interpretation of the kinks (bivalent intersections)? Can
the definition of the square root be extended to incorporate them?
(v) Compute eigenvalues of volume and the action of the Hamiltonian
on higher valence intersections. (vi) Derive transition amplitudes in
a different independent evolution parameter and compare the
results. (vii) Study the physical meaning of the renormalization
constant $Z$ in (\ref{peppo}); is the low energy Newton constant
related to the renormalized or unrenormalized Planck length? (vii)
Study the convergence of the sum (\ref{h}) that defines the node
Hamiltonian and of the sum over intermediate states in second order
transition amplitudes (see eq.(\ref{pippo})); are there
uncontrollable divergences? (vii) Can the classical theory be
reconstructed from the diffeomorphism invariant quantum theory?
(viii) Extend the theory to fermions, along the lines of reference
\cite{fermions}.

\vskip1cm

I thank Bernd Br\"ugmann,  Alain Connes, J\"urgen
Ehlers, Rudolf Haag, Gary Horowitz,  Chris Isham, Ted
Jacobson,  Karel Kucha\v{r},  Ted Newman, Roger Penrose
and Julius Wess, for
various comments, discussions and
difficult questions.  I am particularly indebted to Abhay Ashtekar
for discussions and criticisms, to Ranjeet Tate for important
comments and a careful reading of the manuscript,
and to Lee Smolin: many of the ideas on which this work is based
emerged in discussions with Lee.
This work was partially supported by the NSF grant PHY-9311465.

\end{document}